\documentclass[pra,twoside,showpacs,superscriptaddress,twocolumn,floatfix]{revtex4}

\usepackage{latexsym}
\usepackage[dvips]{graphics}

\begin{document}

\title{Experimental realization of a programmable
       quantum-state discriminator and a phase-covariant
       quantum multimeter}

\author{Jan Soubusta}

\affiliation{Joint Laboratory of Optics of Palack\'{y} University and
     Institute of Physics of Academy of Sciences of the Czech Republic,
     17. listopadu 50A, 772\,00 Olomouc, Czech Republic}
\affiliation{Department of Optics, Palack\'y University,
     17.~listopadu 50, 772\,00 Olomouc, Czech~Republic}

\author{Anton\'{\i}n \v{C}ernoch}

\affiliation{Department of Optics, Palack\'y University,
     17.~listopadu 50, 772\,00 Olomouc, Czech~Republic}

\author{Jarom\'{\i}r Fiur\'{a}\v{s}ek}

\affiliation{QUIC, Ecole Polytechnique, CP 165, Universit\'{e}
     Libre de Bruxelles, 1050 Bruxelles, Belgium}
\affiliation{Department of Optics, Palack\'y University,
     17.~listopadu 50, 772\,00 Olomouc, Czech~Republic}

\author{Miloslav Du\v{s}ek}

\affiliation{Department of Optics, Palack\'y University,
     17.~listopadu 50, 772\,00 Olomouc, Czech~Republic}

\date{\today}

\begin{abstract}
We present an optical implementation of two programmable quantum
measurement devices. The first one serves for unambiguous
discrimination of two nonorthogonal states of a qubit. The
particular pair of states to be discriminated is specified by the
quantum state of a program qubit. The second device can perform von
Neumann measurements on a single qubit in any basis located on the
equator of the Bloch sphere. Again, the basis is selected by the state of
a program qubit. In both cases the data and program qubits
are represented by polarization states of photons. The
experimental apparatus exploits the fact that two Bell states can
be distinguished solely by means of linear optics. The outcome
corresponding to the remaining two Bell states represents an
inconclusive result.
\end{abstract}


\maketitle


\section{Introduction}

Quantum measurements are inevitable parts of all quantum devices
\cite{helst,hol,per,N+Ch}. In many situations, the choice of a
particular measurement depends on the task to be performed. For
instance, in the case of quantum-state discrimination the choice
of the generalized measurement is given by the specific pair of
states that are supposed to be discriminated. Recently universal
(multi-purpose) quantum measurement devices, ``quantum
multimeters'', were introduced and discussed in several papers
\cite{DuBu,FiDuFi,HeDuFiFi,FiDu,Paz}. Their key property is the
possibility to control the choice of the measurement by a quantum
state of the program register, which could be in principle unknown.
The quantum states of a program register corresponding to different
measurements are allowed to be mutually non-orthogonal.

In this paper, we report on the experimental realization of a programmable
quantum-state discriminator (Section \ref{discriminator}) and
a phase-covariant multimeter (Section \ref{multimeter}). Both
programmable detectors involve two qubits, one data qubit and
one program qubit. In our optical implementation,
the qubits are encoded into polarization states of single photons and
the required photon pairs are generated by means
of spontaneous parametric downconversion. The experiment exploits the fact that
a partial Bell measurement on polarization states of two photons can be
accomplished with linear optics, namely a balanced beamsplitter and two
polarizing beamsplitters, followed by photodetectors
\cite{Weinfurter94,Braunstein95,innsbruck,Mattle96,Bouwmeester97}.


\section{Programmable quantum-state discriminator}\label{discriminator}

\subsection{Theory}

A general {\em unknown} quantum state cannot be determined
completely by a measurement performed on a single copy of the
system. But the situation is different if {\em a priori} knowledge
is available \cite{helst,hol,per} -- e.g., if one works only with
states from a certain discrete set. Even quantum states that are
mutually non-orthogonal can be distinguished with a certain
probability provided they are linearly independent (for a review
see Ref.~\cite{Chef}). There are, in fact, two different optimal
strategies \cite{barn-pha}: First, the strategy that determines
the state with the minimum probability for the error
\cite{helst,hol} and, second, unambiguous or error-free
discrimination (the measurement result never wrongly identifies a
state) that allows the possibility of an inconclusive result (with
a minimal probability in the optimal case)
\cite{usd-I,usd-D,usd-P,j+s,c+b}. We will concentrate our
attention to the unambiguous state discrimination. It has been
first investigated by Ivanovic \cite{usd-I} for the case of two
equally probable non-orthogonal states. Peres \cite{usd-P} solved
the problem of discrimination of two states in a formulation with
POVM measurement. Later Jaeger and Shimony \cite{j+s} extended the
solution to arbitrary a priori probabilities. Chefles and Barnett
\cite{c+b} have generalized Peres's solution to an arbitrary
number of equally probable states which are related by a symmetry
transformation. Unambiguous state discrimination was already
realized experimentally. The first experiment, designed for the
discrimination of two linearly polarized states of light, were
done by Huttner \emph{et al.} \cite{discrim}. The interest in the
quantum state discrimination is not only ``academic'' --
unambiguous state discrimination can be used, e.g., as an
efficient attack in quantum cryptography \cite{DJL}.

Let us now suppose that we want to discriminate unambiguously
between two non-orthogonal states. However, we would like to
have a possibility to ``switch'' the apparatus in order to be able
to work with several different pairs of states. The switching should
be realized by preparing a program register of the programmable
discriminator in different program states. The program state thus specifies
which pair of quantum states is discriminated by the device.
This problem was investigated theoretically in Ref.~\cite{DuBu}.

\begin{figure}
  \resizebox{0.9\hsize}{!}{\includegraphics*{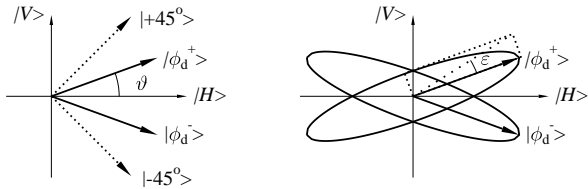}}
  \caption{Polarization of the ``data'' photon: left panel -- linear polarizations,
           right panel -- elliptical polarizations.}
  \label{fi_pm}
\end{figure}

Let us have two (non-orthogonal) input states of a qubit that
should be discriminated:
\begin{equation}
  | \phi^{\pm}_{\mathrm{d}} \rangle =
      a \, | H_{\mathrm{d}} \rangle
  \pm b \, | V_{\mathrm{d}} \rangle,
 \label{data}
\end{equation}
where $| H_{\mathrm{d}} \rangle$ and $|V_{\mathrm{d}} \rangle$
represent two ``logical'' levels of a qubit, in particular
horizontal and vertical linear polarizations of a photon;
subscript d stays for ``data''. Note that these states are
supposed to be symmetrically located around the state
$|H_{\mathrm{d}} \rangle$, as depicted in Fig.~\ref{fi_pm}.
In addition let us have a program
qubit in a state (index p denotes ``program''):
\begin{equation}
  | \phi_{\mathrm{p}} \rangle =
    a' \, | H_{\mathrm{p}} \rangle
  + b' \, | V_{\mathrm{p}} \rangle.
 \label{program}
\end{equation}
Choosing properly the state of a program qubit and performing a suitable
joint measurement on the data and program qubits together one can
unambiguously discriminate any two states of the data qubit that
are in agreement with the program.

Let us suppose that the state of the program qubit is equal to one
of the two states that shall be error-free discriminated. 
In particular, let
\begin{equation}
  | \phi_{\mathrm{p}} \rangle =
   a \, | H_{\mathrm{p}} \rangle
 + b \, | V_{\mathrm{p}} \rangle.
 \label{program!}
\end{equation}
Then the total state of the data and program qubit reads
\begin{eqnarray}
  &| \phi^{\pm}_{\mathrm{d}} \rangle \otimes
   | \phi_{\mathrm{p}} \rangle =& \nonumber \\
  &\sqrt{2} \left[ \displaystyle
  \frac{a^2 \pm b^2}{2} \, |{\Phi^{+}} \rangle +
  \frac{a^2 \mp b^2}{2} \, |{\Phi^{-}} \rangle +
  a b \, |{\Psi^{\pm}} \rangle \right],&
 \label{total}
\end{eqnarray}
where
\begin{eqnarray}
  |\Psi^{\pm} \rangle &=& {1\over\sqrt{2}} \left(
  |H_{\mathrm{d}} \rangle
  |V_{\mathrm{p}} \rangle \pm
  |V_{\mathrm{d}} \rangle
  |H_{\mathrm{p}} \rangle \right),
    \nonumber \\
  |\Phi^{\pm} \rangle &=& {1\over\sqrt{2}} \left(
  |H_{\mathrm{d}} \rangle
  |H_{\mathrm{p}} \rangle \pm
  |V_{\mathrm{d}} \rangle
  |V_{\mathrm{p}} \rangle \right)
\label{BellSt}
\end{eqnarray}
are the Bell states. Clearly, if we were able to detect Bell
states $|\Psi^{+} \rangle$ and $|\Psi^{-} \rangle$ we could
unambiguously discriminate states $| \phi^{+}_{\mathrm{d}}
\rangle$ and $| \phi^{-}_{\mathrm{d}} \rangle$.

The probability of successful discrimination would be
\begin{equation}
  p = 2 |a b|^2 = 2 ( |a|^2 - |a|^4 ).
 \label{probab}
\end{equation}
This probability of success is not optimal in general. As
shown in Refs. \cite{usd-P,usd-I,usd-D} the optimal probability of
successful discrimination $p_{\mathrm{opt}} = 1 - \left|
\langle \phi^{+}_{\mathrm{d}} | \phi^{-}_{\mathrm{d}}
\rangle \right| = 1 - \left| 2|a|^2 - 1 \right|$. In fact,
the probability (\ref{probab}) corresponds to a
``quasi-classical'' discrimination. By a quasi-classical
approach we mean a probabilistic measurement when one
randomly selects
\footnote{With the same probabilities provided that the
    frequencies of the occurrence of the input states are also
    the same.}
the projective measurement in one of two bases that both span the
two-dimensional space containing both non-orthogonal states of
interest (\ref{data}). One basis consists of the state $|
\phi^{+}_{\mathrm{d}} \rangle$ and its orthogonal complement $|
\phi^{+}_{\perp\mathrm{d}} \rangle$. If one finds the result
corresponding to $| \phi^{+}_{\perp\mathrm{d}} \rangle$ he/she can
be sure that the state $| \phi^{+}_{\mathrm{d}} \rangle$ was not
present. Analogously, the other basis consists of the state $|
\phi^{-}_{\mathrm{d}} \rangle$ and its orthogonal complement.

To get a higher probability of success one would have to make a
more sophisticated measurement than a simple Bell-state analysis
\cite{DuBu}. However, there are two important points concerning
the proposed discrimination procedure: First, a partial
measurement in the Bell basis can be easily realized by linear
optics. Namely, states $|\Psi^{+} \rangle$ and $|\Psi^{-} \rangle$
can be distinguished (the rest of the Bell states corresponds to
an inconclusive result anyway). Second, our aim is not to maximize
the success probability but to demonstrate experimentally the
possibility to control the discrimination process by the
\emph{quantum state} of a program qubit. Even to set the bases for
the mentioned quasi-classical discrimination correctly, one needs
an infinite number of bits of classical information. Our
procedure requires only \emph{a single qubit} for the same job. This
fact reveals the \emph{quantum nature} of the programming. Even
non-orthogonal states of the quantum qubit carry ``useful
information'' for the discrimination process.


\subsection{Experiment}

\begin{figure}
  \begin{center}
    \smallskip
  \resizebox{0.8\hsize}{!}{\includegraphics*{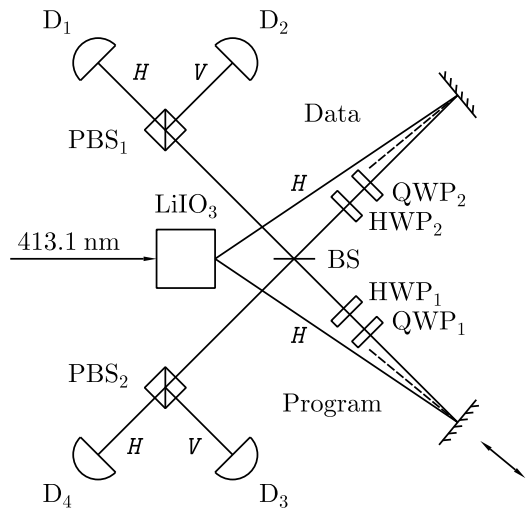}}
    \smallskip
  \end{center}
  \caption{Experimental setup (BS is a beamsplitter,
  PBS polarization beamsplitters, D denotes detectors,
  and HWP and QWP half- and quarter-wave plates, respectively).}
  \label{schema}
\end{figure}

The scheme of our experimental setup is shown in
Fig.~\ref{schema}. A krypton-ion cw laser (413.1~nm, 95 mW)
is used to pump a 10-mm-long LiIO$_3$ nonlinear crystal cut
for degenerate type-I parametric downconversion. We
exploit the fact that the pairs of photons generated by
spontaneous parametric downconversion (SPDC) manifest
tight time correlations (i.e., very exact coincidences of
detection instants). In our setup the photons produced by
SPDC have horizontal linear polarizations. Different
polarization states are prepared by means of half-wave and
quarter-wave plates (HWP, QWP). The polarization of the
signal photon represents the data qubit whilst the
polarization of the idler photon serves as the program qubit.
The two photons impinge on two input ports of a 50/50 beamsplitter
(BS, non-polarizing cube beamsplitter). A scanning mirror is
used in one interferometer arm in order to balance the
length of both arms, as indicated by an arrow in
Fig.~\ref{schema}.
The photons reflected and transmitted by BS pass polarizing
beamsplitters (PBS, polarizing cube beamsplitter) to distinguish horizontal
and vertical polarizations. Finally, the beams are filtered by
cut-off filters and circular apertures, and coupled into
multi-mode optical fibers by lenses. Detectors D$_1$,\ldots,D$_4$
are Perkin-Elmer single-photon counting modules (employing silicon
avalanche photodiodes with quantum efficiency $\eta\approx 50\,$\%
and dark counts about 100\,s$^{-1}$). The signals from detectors
are processed by our home-made 4-input coincidence module. With this
setup, visibilities of Hong-Ou-Mandel dip \cite{Mandel} exceeding 92\,\% were
reached for vertically polarized photons (for other polarizations
visibilities were slightly lower). Higher visibilities could not
be reached due to the fact that the splitting ratios of the
coupler were not exactly 50/50 and due to the imperfections of the
wave plates and polarization distortions on mirrors.


As already mentioned our experiment is based on the
possibility to detect the two particular Bell states in the
Hilbert space of the polarization states of two photons.
This task can be done by means of passive linear optical
elements, namely by a beamsplitter and two polarization
beamsplitters as shown in Fig.~\ref{schema}
\cite{Weinfurter94,Braunstein95,innsbruck,Mattle96,Bouwmeester97}.

The simplest theoretical model of the beamsplitter leads to
the conclusion that if one fetches Bell states at the
inputs the only one of them that results in a coincident
detection at two different outputs of the beamsplitter is
the singlet state $|\Psi^{-} \rangle$. However, in case of
the ``real'' beam-splitting cube one must take into account
that the two photons strike upon a beamsplitter in opposite
directions. So, if the mutual phase of the vertical
components of the electric-field vectors at the interface
plane (we mean the vectors corresponding to the fields
coming from the two opposite inputs) is $\varphi$ the
mutual phase of the horizontal components is $\varphi +
180^\circ$ just for \emph{geometrical} reasons. It means,
if one prepares both photons in the \emph{same} linear
polarizations tilted by $45^\circ$ with respect to the
plane of incidence the vectors of electric field at the
interface plane will oscillate in mutually
\emph{perpendicular} directions. Consequently, now it is
the triplet state $|\Psi^{+} \rangle$ that leads to a
coincident detection at different outputs.

So, if detectors D$_1$ and D$_3$ or D$_2$ and D$_4$ click
together in our setup the state $|\Psi^{+} \rangle$ is
detected (at the input of the beamsplitter)
--- this corresponds to the recognition of state $|
\phi^{+}_{\mathrm{d}} \rangle$.
If detectors D$_1$ and D$_2$ or D$_3$ and D$_4$
click in coincidence the state $|\Psi^{-} \rangle$ was
``present'' in the input --- so, $| \phi^{-}_{\mathrm{d}}
\rangle$ is detected.
If both photons enter the same detector either $|\Phi^{+}
\rangle$ or $|\Phi^{-} \rangle$ was ``present'' in the
input --- this represents the inconclusive result of the
discrimination.

Each data point at presented plots has been derived from 10
one-second measurement periods.
The accuracy of polarization-angle settings was better than
$\pm 1^\circ$.

\subsection{Results}


\begin{figure}
  \resizebox{\hsize}{!}{\includegraphics*{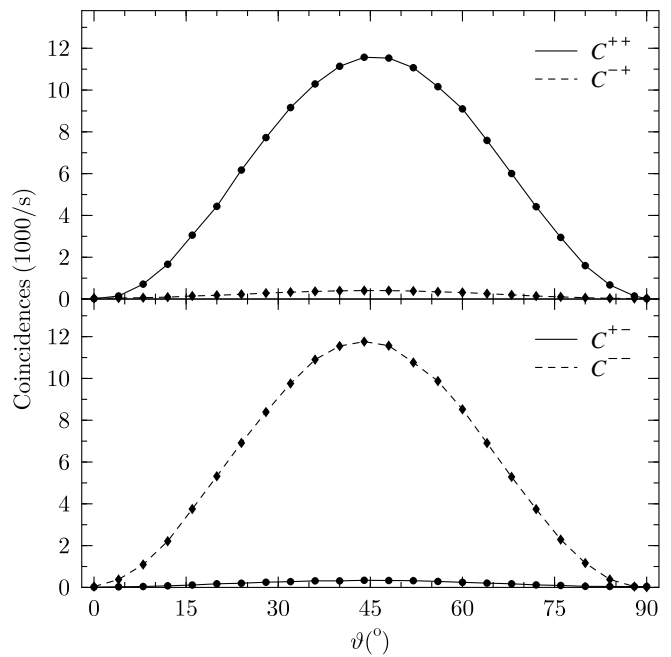}}
  \caption{Detection rates of $|\Psi^{+} \rangle$ (dots, solid line) and
           $|\Psi^{-} \rangle$ (diamonds, broken line)
           for linearly polarized input states. The upper part corresponds
           to the situation when the data and program states coincide,
           $|\phi^{+}_{\mathrm{d}} \rangle \otimes |\phi_{\mathrm{p}} \rangle$,
           the lower part to the situation when they are different,
           $|\phi^{-}_{\mathrm{d}} \rangle \otimes |\phi_{\mathrm{p}} \rangle$.}
  \label{lin_coinc}
\end{figure}


\begin{figure}
  \resizebox{\hsize}{!}{\includegraphics*{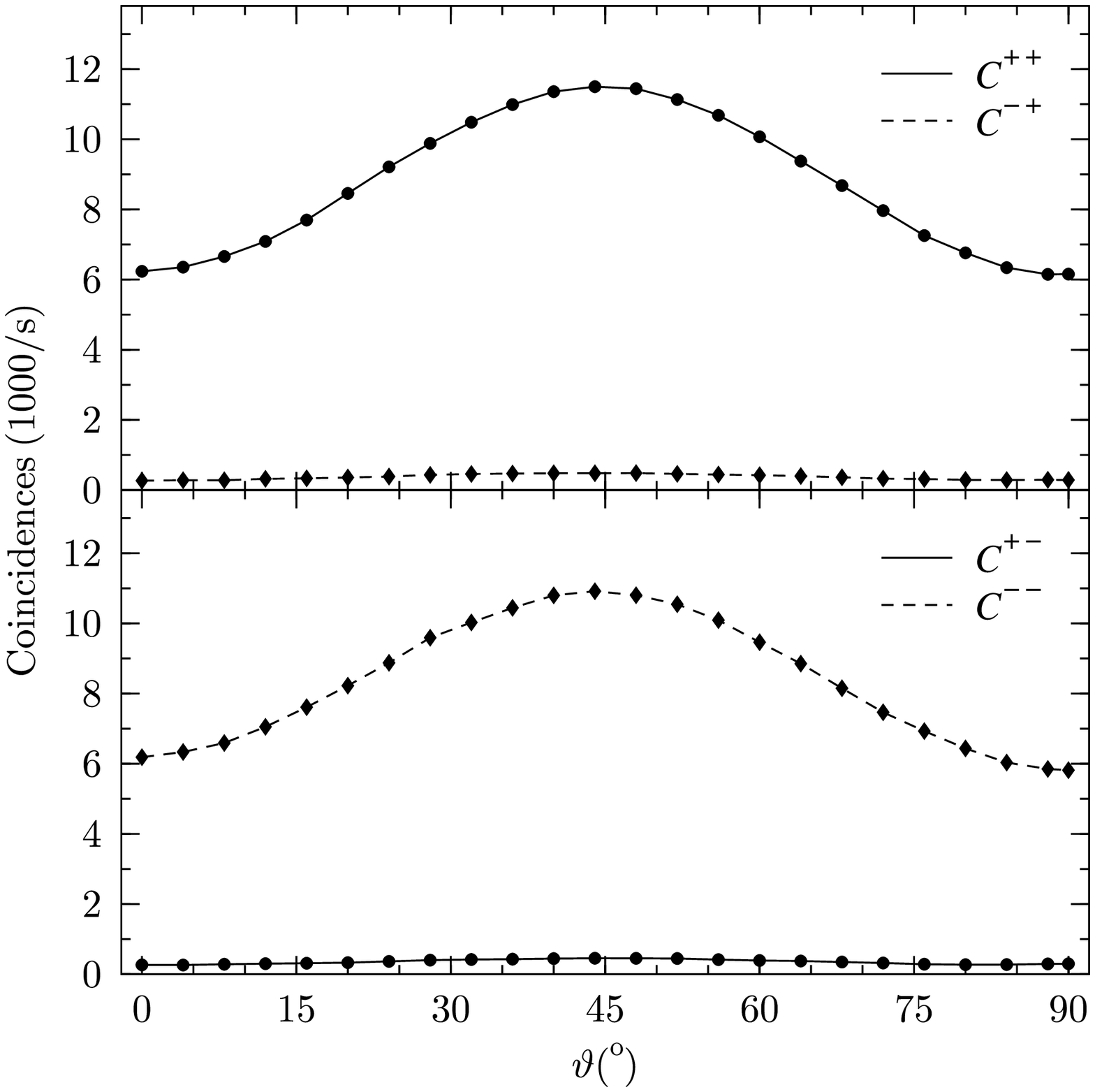}}
  \caption{Detection rates of $|\Psi^{+} \rangle$ (dots, solid line) and
           $|\Psi^{-} \rangle$ (diamonds, broken line)
           for elliptically polarized input states, $\varepsilon=24^\circ$.
           The upper part corresponds to the situation when the data
           and program states coincide,
           $|\phi^{+}_{\mathrm{d}} \rangle \otimes |\phi_{\mathrm{p}} \rangle$,
           the lower part to the situation when they are different,
           $|\phi^{-}_{\mathrm{d}} \rangle \otimes |\phi_{\mathrm{p}} \rangle$.}
  \label{ellip_coinc}
\end{figure}


We tried to discriminate a broad variety of pairs of elliptically
polarized states
\begin{equation}
 | \phi^{\pm}_{\mathrm{d}} \rangle =
      (x \cos \vartheta + i y \sin \vartheta) \, | H_{\mathrm{d}} \rangle
  \pm (x \sin \vartheta - i y \cos \vartheta) \, | V_{\mathrm{d}} \rangle,
 \label{elipt}
\end{equation}
where $x$ and $y$ are the half-axes of the polarization ellipse
and $\vartheta$ is the angle between horizontal axis and the direction of
$x$-half-axis (see Fig.~\ref{fi_pm}). The ellipticity
$\tan(\varepsilon)=y/x$. We started with
linear polarizations ($\varepsilon = 0$) and then proceeded
with elliptical polarizations with three different
ellipticities (namely for $\varepsilon=12^\circ, 24^\circ$,
and $36^\circ$). The angle $\vartheta$ was, in all the
cases, scanned from $0^\circ$ to $90^\circ$ with the step
of $4^\circ$. Elliptical polarizations (corresponding to states $|
\phi^{\pm}_{\mathrm{d}} \rangle$) were produced from
horizontal ones by means of a quarter-wave plate rotated by
angle $\alpha=\pm\varepsilon$ with respect to the horizontal
axis. Then the direction of the $x$-half-axis of the
polarization ellipse was tilted by a half-wave plate
rotated by $\beta=\pm(\varepsilon+\vartheta)/2$.

Let us recall that the theoretical probability of successful
discrimination of states (\ref{elipt}) is
\begin{equation}
  p = 2 ( |a|^2 - |a|^4 ), \mbox{~where~~}
  |a|^2 = x^2 \cos^2 \vartheta +y^2 \sin^2 \vartheta.
 \label{p_teor}
\end{equation}

Figure~\ref{lin_coinc} shows the coincidence rates measured for
linear polarization states. The upper part shows the measurement
of Bell states $|\Psi^{+} \rangle$ and $|\Psi^{-} \rangle$ when
the data and program states coincide (i.e., the input state is
$|\phi^{+}_{\mathrm{d}} \rangle \otimes |\phi_{\mathrm{p}}
\rangle$), the lower part shows the same measurement in case when
they are different (i.e., the input state is
$|\phi^{-}_{\mathrm{d}} \rangle \otimes |\phi_{\mathrm{p}}
\rangle$). Statistical errors are smaller than the symbols of points.
Graphs in Fig.~\ref{ellip_coinc} illustrate the same
measurement but with pairs of elliptically polarized states
($\varepsilon=24^\circ$, i.e., $\tan(\varepsilon)=0.45$).

Throughout the whole paper,
$C^{++}$ denotes the detection rate of $|\Psi^{+} \rangle$
when the input was in the state
$|\phi^{+}_{\mathrm{d}} \rangle \otimes |\phi_{\mathrm{p}} \rangle$,
$C^{-+}$ denotes the detection rate of $|\Psi^{-} \rangle$
when the input was in the state
$|\phi^{+}_{\mathrm{d}} \rangle \otimes |\phi_{\mathrm{p}} \rangle$,
$C^{+-}$ denotes the detection rate of $|\Psi^{+} \rangle$
when the input was in the state
$|\phi^{-}_{\mathrm{d}} \rangle \otimes |\phi_{\mathrm{p}} \rangle$,
and
$C^{--}$ denotes the detection rates of $|\Psi^{-} \rangle$
when the input was in the state
$|\phi^{-}_{\mathrm{d}} \rangle \otimes |\phi_{\mathrm{p}} \rangle$.

The relative (with respect to all conclusive results) error
rates -- i.e., fractions of events when we get $|\Psi^{+}
\rangle$ instead of $|\Psi^{-} \rangle$ or viceversa -- are
about $5\,$\%. This value is approximately the same for all
measured polarizations. Ideally the error rate should be
zero but due to technical imperfections of the setup it
gets this small nonzero value.

\begin{figure}
  \resizebox{\hsize}{!}{\rotatebox{-90}{\includegraphics*{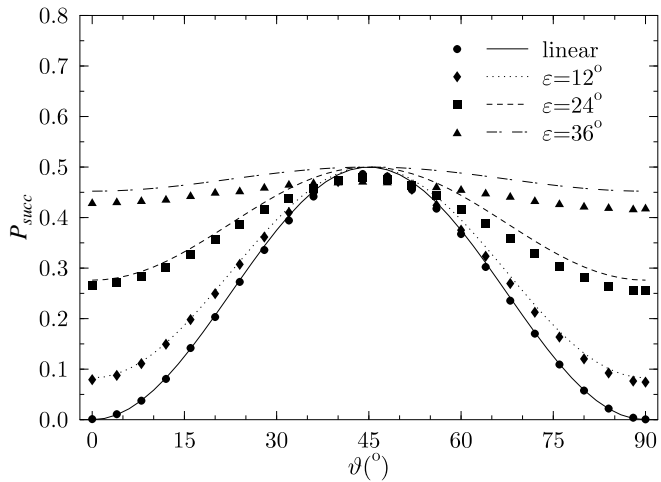}}}
  \caption{Probability of successful operation as a function
           of angle $\vartheta$. Symbols denote experimental data,
           lines represent theoretical predictions (\ref{p_teor}).
           The first curve corresponds to linearly
           polarized photons, the others, in
           sequence, to elliptically polarized photons with
           $\varepsilon=12^\circ, 24^\circ$, and $36^\circ$.}
  \label{p_succ}
\end{figure}

The main parameter -- the probability of success -- is
plotted in Fig.~\ref{p_succ} for four different
ellipticities. Symbols denote experimental data, lines
represent theoretical predictions (\ref{p_teor}). The first
curve corresponds to linearly polarized photons (acting as
data and program qubits), the others, in sequence, to
elliptically polarized photons with $\varepsilon=12^\circ,
24^\circ$, and $36^\circ$.
As can be seen the data are in a good agreement with
the theory.

\begin{figure}
  \resizebox{\hsize}{!}{\rotatebox{-90}{\includegraphics*{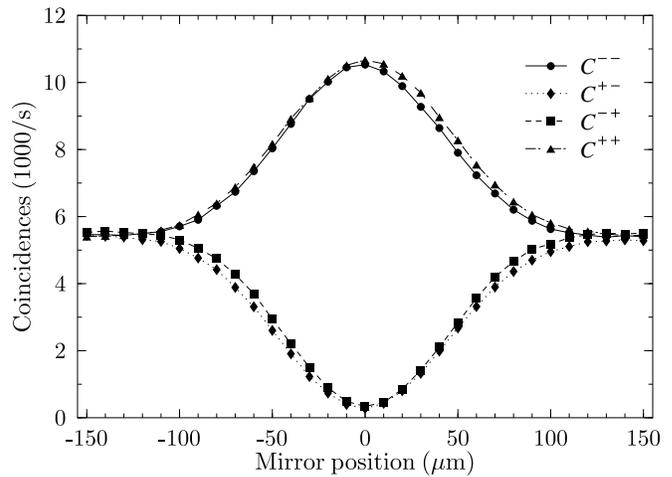}}}
  \caption{Detection rates as functions of the mirror position.
  Here $C^{++}$ is a coincidence rate at D$_{1}$ and D$_{3}$ or
  D$_{2}$ and D$_{4}$ when the input polarizations were $45^\circ,45^\circ$.
  The quantity $C^{+-}$ is a coincidence rate at the same detectors for
  input polarizations $-45^\circ,45^\circ$. Analogously, $C^{-+}$ and $C^{--}$
  are coincidence rates at D$_{1}$ and D$_{2}$ or D$_{3}$ and D$_{4}$.}
  \label{dip}
\end{figure}

The probability of success is calculated as an average of
normalized detection rates corresponding to successful
events (i.e., correct conclusive measurement results) for two possible
input states (that are in correspondence with the program
state). The averaging takes into account that both the
inputs $|\phi^{+}_{\mathrm{d}} \rangle \otimes
|\phi_{\mathrm{p}} \rangle$ and $|\phi^{-}_{\mathrm{d}}
\rangle \otimes |\phi_{\mathrm{p}} \rangle$ appear with the
same probability $1/2$. The ``normalization'' means that
the detection rate is divided by the pair-generation rate,
i.e., by the number of all measurement events per time
unit. This quantity is obtained from the coincidence measurement
on the shoulder $150\,\mu$m out of the dip (see Fig.~\ref{dip}).
In this situation the two one-photon packets generated by
downconversion no longer overlap at the beamsplitter and each
of them randomly ``decides'' whether to go through or to be
reflected.

As the splitting ratio of the used beamsplitter is slightly
different for vertical and horizontal linear polarizations
the normalization measurements were done with $45^\circ$
linear polarization states at the inputs. For the set of
main measurements (i.e., Bell measurements with \emph{balanced}
arms of Hong-Ou-Mandel interferometer)
with the states $|\phi^{+}_{\mathrm{d}}
\rangle \otimes |\phi_{\mathrm{p}} \rangle$ the input
polarization of the normalization measurement were
$45^\circ$, $45^\circ$, for the set of main measurements with
the states $|\phi^{-}_{\mathrm{d}} \rangle \otimes
|\phi_{\mathrm{p}} \rangle$ they were $-45^\circ$,
$45^\circ$.

Because our setup and detection electronics are tailored to
the measurement of the Bell states $|\Psi^{+} \rangle$ and
$|\Psi^{-} \rangle$ we can measure only the events when
detectors D$_1$ and D$_3$ or D$_2$ and D$_4$ click together
or when detectors D$_1$ and D$_2$ or D$_3$ and D$_4$ click
in coincidence. For the given $45^\circ$ polarizations both
these two rates outside the dip should be equal to $1/4$ of
the pair-generation rate. However, there could be
deviations from this $1/4$ due to the non-ideal splitting
ratio of the beamsplitter. Nevertheless, the sum of these two
quantities is immune against this imperfection and
proportional to $1/2$ of the pair-generation rate.

Thus the probability of success is calculated as follows
\begin{equation}
 P_{succ}=\frac 12 \left[
  \frac {C^{++}}{2 (C^{++}_{\rm sh} + C^{-+}_{\rm sh})}
+ \frac {C^{--}}{2 (C^{--}_{\rm sh} + C^{+-}_{\rm sh})}
\right],
 \label{P_succ}
\end{equation}
where $C_{\rm sh}$ are the above mentioned detection rates
measured outside the dip; in particular, $C^{++}_{\rm sh}$
denotes the coincidence rate of D$_1$ and D$_3$ or D$_2$
and D$_4$ with $45^\circ$, $45^\circ$ polarizations at the
inputs, etc. These quantities were always measured just
before, or just after the corresponding set of main measurements
in order to minimize the influence of the long-term fluctuations
(like a laser-power drift etc.).
The errors derived from the statistical fluctuations of
individual detection rates are smaller than the symbols of
points in Fig.~\ref{p_succ}. The systematic error (the
shift of all the data values) is due to non-unit
visibility.


\section{Phase-covariant quantum multimeter}\label{multimeter}

\subsection{Theory}

Now we will consider a different kind of quantum multimeter.
Namely such one that can perform von Neumann measurements on a
single qubit in any basis $\{|\psi_{+}\rangle,|\psi_{-}\rangle\}$
located on the equator of the Bloch sphere,
\begin{equation}
|\psi_{\pm}(\phi)\rangle=\frac{1}{\sqrt{2}}(|0\rangle\pm
e^{i\phi}|1\rangle),
\label{pcbasis}
\end{equation}
where $\phi\in[0,2\pi)$ is arbitrary. The particular measurement
basis is selected by a quantum state of a program register.

It is impossible to perfectly encode such projective
measurements into states in finite-dimensional Hilbert
space \cite{DuBu,FiDu}. Nevertheless, it is possible to encode POVMs
that represent, in a certain sense, the best approximation
of the required projective measurements. A specific way of
approximation of a projective measurement is a
``probabilistic'' measurement that allows for some
inconclusive results. In this case, instead of a
two-component projective measurement one has a
three-component POVM and the third outcome corresponds to
an inconclusive result. The optimal multimeter should
minimize the error rate at the first two outcomes for a fixed
fraction of inconclusive results. As a limit case it is
possible to get an error-free operation. Such a multimeter
performs the exact projective measurements but with the
probability of success lower than one.

The optimal phase-covariant multimeters  for the program states
$|\Psi\rangle_{\mathrm{p}}$ consisting of $N$ copies of the
basis state $|\psi_{+}\rangle$,
$|\Psi\rangle_{\mathrm{p}}=|\psi_+\rangle^{\otimes N}$
\footnote{Since the state $|\psi_{-}\rangle$ can be
obtained form $|\psi_{+}\rangle$ via unitary
transformation, $|\psi_{-}\rangle=\sigma_z|\psi_{+}\rangle$
where $\sigma_z$ denotes the Pauli matrix, all the programs
of the form $|\psi_{+}\rangle^{\otimes j}
|\psi_{-}\rangle^{\otimes N-j}$ are equivalent to the
program $|\psi_{+}\rangle^{N}$.}, were determined in \cite{FiDu}.
For the simplest case $N=1$ that corresponds to a one-qubit
program the optimal (fixed) POVM acting on data and program
qubits together reads:
\begin{eqnarray}
  \displaystyle
  \Pi_{\pm} &=& |\Psi^{\pm} \rangle \langle \Psi^{\pm}|
  +\frac{1-\eta}{2}( |\Phi^{+} \rangle \langle \Phi^{+}|
                    +|\Phi^{-} \rangle \langle \Phi^{-}|),
\nonumber \\
  \displaystyle
  \Pi_{?} &=& \eta ( |\Phi^{+} \rangle \langle \Phi^{+}|
                    +|\Phi^{-} \rangle \langle \Phi^{-}|),
 \label{POVM}
\end{eqnarray}
where $\eta \in [0,1]$, and
\begin{eqnarray}
  |\Psi^{\pm} \rangle &=& \frac{1}{\sqrt{2}} \left(
     |0_{\mathrm{d}} \rangle |1_{\mathrm{p}} \rangle \pm
     |1_{\mathrm{d}} \rangle |0_{\mathrm{p}} \rangle \right),
\nonumber \\
  |\Phi^{\pm} \rangle &=& \frac{1}{\sqrt{2}} \left(
     |0_{\mathrm{d}} \rangle |0_{\mathrm{p}} \rangle \pm
     |1_{\mathrm{d}} \rangle |1_{\mathrm{p}} \rangle \right).
\end{eqnarray}

The fidelity of the multimeter can be defined as the probability of
a correct measurement outcome in the case of a conclusive result,
assuming the data register is prepared in the basis state $|\psi_+\rangle$
or $|\psi_-\rangle$ with probability $1/2$ each.
The mean fidelity is then  obtained by averaging the fidelity over all
measurement bases, i.e. over the phase $\phi\in[0, 2\pi)$. The average fidelity
of the phase-covariant multimeter depends on the probability of inconclusive
result in the following way,
\[
F = \frac{3 - 2 P_I}{4(1-P_I)}.
\]

The effective POVM on the data qubit only (this POVM depends on
the program state) is given by
\begin{eqnarray*}
\pi_{\pm}&=&(1-P_{I})[ F \, |\psi_{\pm} (\phi)
\rangle\langle \psi_{\pm} (\phi)|  \\
& & + \, (1-F) \, |\psi_{\mp} (\phi) \rangle\langle
\psi_{\mp} (\phi) | ],
\\
\pi_{?}&=&P_{I}\openone.
\end{eqnarray*}

For $\eta = 0$ we have an ambiguous (error-prone) operation
with no inconclusive results ($P_I=0, F=3/4$) whilst for
$\eta = 1$ we get an unambiguous (error-free) but
probabilistic measurement device (from time to time we
obtain an inconclusive result; $P_I=1/2, F=1$)
\footnote{Notice that an unambiguous phase-covariant
multimeter with a single-qubit program represents, in a certain sense,
a limit case of the above-discussed programmable discriminator
corresponding to the discrimination of two orthogonal states.}.

Clearly, when $\eta = 1$ then POVM (\ref{POVM}) is just a
projective measurement on the Bell states $|\Psi^{+}
\rangle$, $|\Psi^{-} \rangle$, and on the rest of the four
dimensional Hilbert space (that corresponds to inconclusive
results). If $\eta < 1$ we just ``reinterpret'' some
inconclusive results as ``conclusive'' ones. I.e., we will
treat randomly selected (with probability $1-\eta$)
inconclusive results as results ``$+$'' or ``$-$'' (at
random). Therefore, it is quite enough to test
experimentally only the unambiguous version ($\eta = 1$) of
our phase-covariant quantum multimeter as all the other
variants ($0 \le \eta < 1$) can be obtained manipulating
the measured data only.

\subsection{Experiment}

The simplest ($N=1$) phase-covariant multimeter can be
experimentally implemented using the same setup as for the
programmable quantum-state discriminator (see
Fig.~\ref{schema}). The ``logical values'' 0 and 1 of
qubits will be represented by horizontal, $|H\rangle$, and
vertical, $|V\rangle$, linear-polarization states of
photons, respectively.

The states $|\psi_{+}(\phi)\rangle$ (\ref{pcbasis}) are
prepared from states $|H\rangle$ by two wave plates. The
first one is a quarter-wave plate rotated by
$\alpha=-\phi/2$ (with respect to the horizontal axis) and
the second one is a half-wave plate rotated by
$\beta=(90^\circ-\phi)/4$. The states
$|\psi_{-}(\phi)\rangle$ are prepared with the wave plates
rotated by angles $-\alpha$ and $-\beta$. This corresponds
to the shift of $\phi$ by $180^\circ$.
Detection of $|\Psi^{+} \rangle$ corresponds to the measurement
result connected with the basis state $|\psi_{+}(\phi)\rangle$,
detection of $|\Psi^{-} \rangle$ corresponds to the measurement
result connected with the basis state $|\psi_{-}(\phi)\rangle$.
Everything else means an inconclusive result.

\subsection{Results}

We have measured coincidence rates corresponding to
detections of $|\Psi^{\pm} \rangle$ and calculated
probabilities of inconclusive results for phases $\phi$
from $-90^\circ$ to $+90^\circ$ with step $8^\circ$. The
inconclusive-result rate is a ``complement'' to the rate of
``conclusive'' results -- i.e., both the correct and
erroneous ones. Thus
\begin{equation}
 P_{I}= 1 - \frac 12 \left[
  \frac {C^{++} + C^{-+}}{2 (C^{++}_{\rm sh} + C^{-+}_{\rm sh})}
+ \frac {C^{--} + C^{+-}}{2 (C^{--}_{\rm sh} + C^{+-}_{\rm
sh})} \right].
 \label{P_inc}
\end{equation}
The notation is the same as above.

\begin{figure}
  \resizebox{\hsize}{!}{\rotatebox{-90}{\includegraphics*{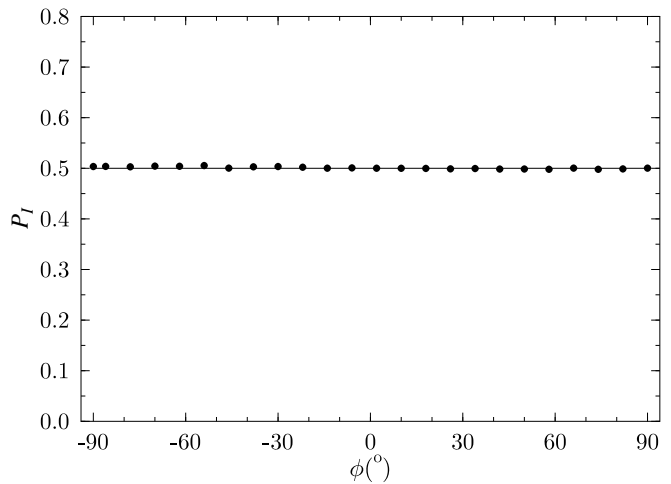}}}
  \caption{Probability of inconclusive results for unambiguous
           operation as a function of phase $\phi$. Symbols denote
           experimental data, solid line
           represents the theoretical prediction $P_I=1/2$.}
  \label{ph_cov_PI}
\end{figure}

Graph in Fig.~\ref{ph_cov_PI} shows the dependence of the
probability of inconclusive results (for an unambiguous operation) on
phase $\phi$ for input data states $|\psi_{+}\rangle$ and
$|\psi_{-}\rangle$ (that coincide with or are orthogonal to the program
states, respectively). Symbols denote experimental data, solid line
represents the theoretical prediction. The derived errors are smaller
than the symbols of points. The experimental data fit well
with the calculated value of $P_I=1/2$. As expected, the
probability of inconclusive results does not depend on the
measurement basis determined by the phase $\phi$.

The relative error rate (with respect to all conclusive
results) --- i.e., a fraction of events when
we get $|\Psi^{+} \rangle$ instead of $|\Psi^{-} \rangle$
or viceversa --- is about $3\,$\%. Ideally, in the considered case of
the unambiguous operation, the error rate should be zero, but
due to technical imperfections of the setup it gets non-zero values.


\section{Conclusions}

Our experimental results clearly illustrate that the measurement
on the data qubit can be quite efficiently controlled by the
quantum state of the program qubit (program register). In
particular, we emphasize that a classical setting of the angle
between the states that shall be unambiguously discriminated and a
classical description of the measurement basis in case of
projective phase-covariant measurement would require infinitely
many bits of classical information, while only one quantum bit
suffices in the present case to obtain an \emph{error-free}
(although probabilistic) operation.


\begin{acknowledgments}

The authors would like to thank to Martin Hendrych and Miroslav Je\v{z}ek
for their help and advises in the preparatory stage of the
experiment. This research was supported under the project
LN00A015 of the Ministry of Education of the Czech
Republic. JF also acknowledges support from the EU under the project CHIC
(IST-2001-32150).

\end{acknowledgments}


\end{document}